\documentclass[sigconf]{acmart}
\usepackage{multicol}
\usepackage{multirow}
\usepackage{hyperref}

\AtBeginDocument{%
  }

\acmConference[ACM Multimedia 2025]{Make sure to enter the correct
  conference title from your rights confirmation email}{October 27--31,
  2025}{Dublin, Ireland}
\acmISBN{978-1-4503-XXXX-X/2018/06}

\renewcommand\footnotetextcopyrightpermission[1]{}

\begin{document}
\title{MusFlow: Multimodal Music Generation via Conditional Flow Matching}

\author{Jiahao Song}
\email{202322208028@email.sxu.edu.cn}
\affiliation{
	\institution{School of Mathematics and Statistics, Shanxi University}
	\city{Taiyuan, Shanxi}
	\country{China}
}

\author{Yuzhao Wang}
\email{wangyuzhao@sxu.edu.cn}
\affiliation{
	\institution{School of Mathematics and Statistics, Shanxi University}
	\city{Taiyuan, Shanxi}
	\country{China}
}



\begin{abstract}
Music generation aims to create music segments that align with human aesthetics based on diverse conditional information. Despite advancements in generating music from specific textual descriptions (e.g., style, genre, instruments), the practical application is still hindered by ordinary users’ limited expertise or time to write accurate prompts. To bridge this application gap, this paper introduces MusFlow, a novel multimodal music generation model using Conditional Flow Matching. We employ multiple Multi-Layer Perceptrons (MLPs) to align multimodal conditional information into the audio's CLAP embedding space. Conditional flow matching is trained to reconstruct the compressed Mel-spectrogram in the pretrained VAE latent space guided by aligned feature embedding. MusFlow can generate music from images, story texts, and music captions. To collect data for model training, inspired by multi-agent collaboration, we construct an intelligent data annotation workflow centered around a fine-tuned Qwen2-VL model. Using this workflow, we build a new multimodal music dataset, MMusSet, with each sample containing a quadruple of image, story text, music caption, and music piece. We conduct four sets of experiments: image-to-music, story-to-music, caption-to-music, and multimodal music generation. Experimental results demonstrate that MusFlow can generate high-quality music pieces whether the input conditions are unimodal or multimodal. We hope this work can advance the application of music generation in multimedia field, making music creation more accessible. Our generated samples, code and dataset are available at \href{https://anonymous22356.github.io/musflow.github.io/}{musflow.github.io}.
\end{abstract}

\ccsdesc[500]{Information systems~Multimedia content creation}
\ccsdesc[500]{Applied computing~Sound and music computing}

\keywords{Music Generation, Multimedia Creation, Flow Matching, Multi-Agent Workflow}

\maketitle

\begin{figure}[h]
  \centering
  \includegraphics[width=\linewidth]{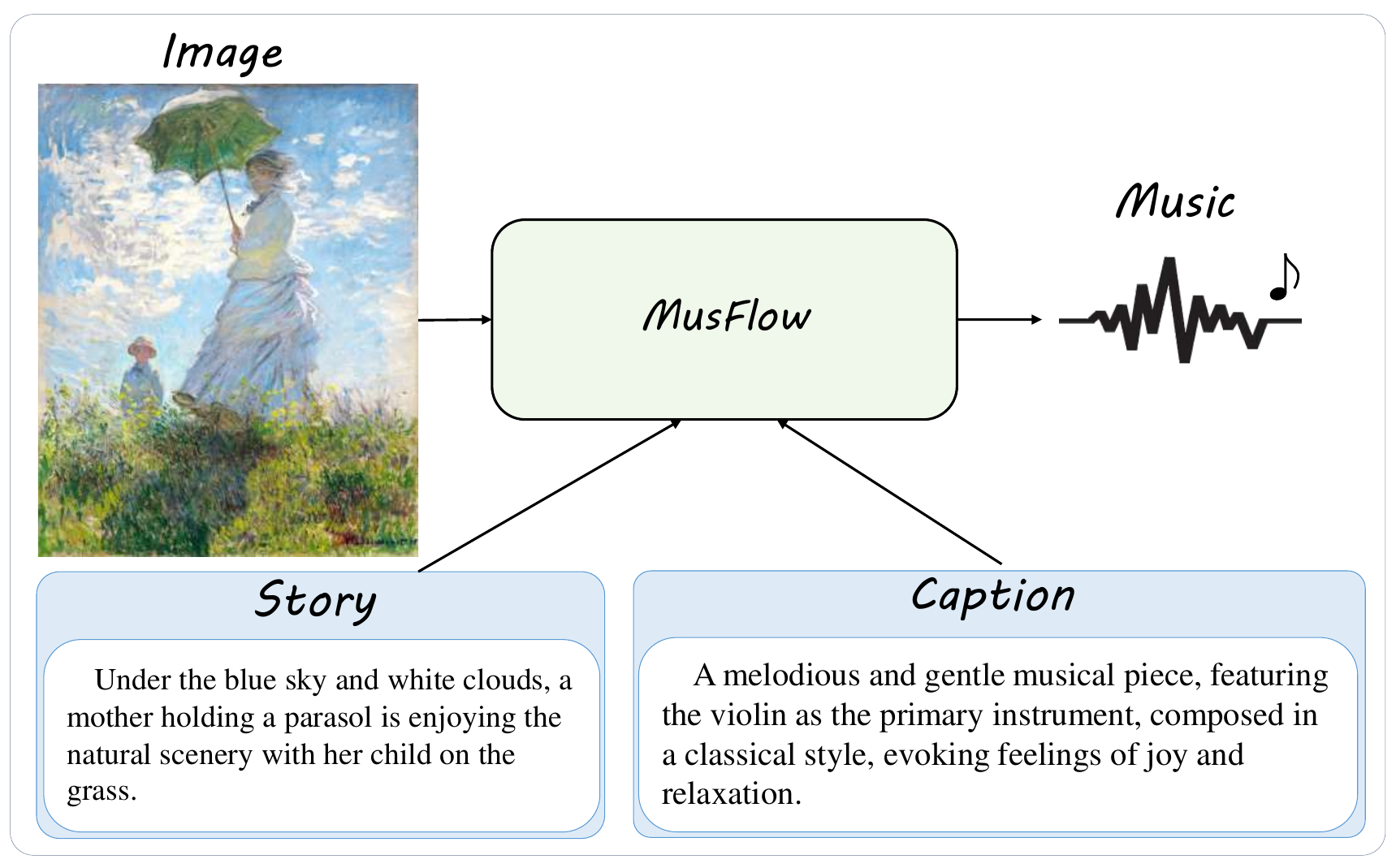}
  \caption{Multimodal music generation by our proposed
MusFlow model.}
  \label{Figure 1}
\end{figure}

\section{Introduction}
Music has long been regarded as a  medium for human emotions, playing a vital role in modern society, especially in multimedia applications \cite{tan2013psychology}. Music generation aims to create musical segments that align with users' aesthetics and requirements by processing conditional information, thereby enriching creative methods in music composition \cite{wang2024review}. 

With the rapid advancement of deep learning technologies, the field of music generation has seen remarkable progress in recent years \cite{briot2020deep}.  Inspired by the success of Transformer and Diffusion models in text and image generation \cite{vaswani2017attention, ho2020denoising, rombach2022high}, researchers have increasingly applied these powerful generative AI techniques to music generation, achieving remarkable progress—especially in text-to-music generation \cite{agostinelli2023musiclm, NEURIPS2023musicgen, chen2024musicldm}. These models allow users to generate music by providing descriptive text that includes musical attributes such as genre, rhythm, instrumentation, and emotion, greatly facilitating the process of music composition.

Despite the significant progress, the application of current music generation models still faces limitations. One major challenge lies in the fact that these models require users to provide music-specific descriptive texts, which demands a certain level of musical expertise. Although music generation models have lowered the barrier for music composition, users are still required to think like composers by conceptualizing the musical attributes, while the model acts more like a performer, playing the specified music. However, in many cases, users prefer music models to possess emotional understanding and creative abilities akin to a composer, generating suitable musical pieces based on contextual information such as images or story texts that lack explicit musical features. For example, in the film industry, producers need to create background music for specific scenes and events in scripts. Similarly, in the audiobook and podcast domains, hosts often need background music that matches a particular storyline. In such cases, models must exhibit artistic comprehension, enabling them to autonomously compose background music that aligns with multimodal information. Multimodal music generation is still underexplored. Existing models that generate music from multimodal conditions often depend on large pretrained language models to handle multimodal inputs \cite{liu2024audioldm, liu2024mumu}, making them computationally expensive and less practical.

As illustrated in Fig. \ref{Figure 1}, in this paper, we propose a novel multimodal music generation model, MusFlow, capable of generating music based on images, story texts, and music captions. Inspired by the success of Flow Matching in the field of image generation \cite{lipman2023flow, liu2023flow}, we apply the Conditional Flow Matching method to the domain of multimodal music generation. To handle multimodal information, we employ MLP Adapters \cite{NEURIPS2023llava} to map multimodal conditional embeddings to the CLAP \cite{laionclap2023} audio feature space for feature alignment. This alignment transforms the conditional information into feature embeddings with musical semantics. Subsequently, we use a pretrained VAE \cite{Kingma2014} to compress the Mel-spectrograms of audio, and conditional flow matching to reconstruct the Mel-spectrograms in the latent space guided by the aligned conditional information. Finally, a pretrained HiFi-GAN model generates the waveform of the music clips from the decoded Mel-spectrograms. Leveraging the excellent sampling efficiency of Flow Matching, this approach enables fast multimodal music generation.

The main limitation of earlier music generation models is the lack of rich, multimodal data. Existing datasets are typically built by having professionals write descriptions for music clips \cite{agostinelli2023musiclm}, resulting in text-only annotations without accompanying context like videos, images, or story texts. To address the scarcity of multimodal music data and the high costs associated with manual annotation, this paper proposes an intelligent multimodal music data annotation workflow based on the Multi-Agent paradigm \cite{hong2024metagpt}. Leveraging pretrained BLIP \cite{li2022blip}, CLIP \cite{radford2021learning}, and CLAP \cite{laionclap2023}, as well as a fine-tuned Qwen2-VL-7B model \cite{wang2024qwen2}, we achieve the autonomous matching and filtering between public image datasets and music datasets. Using this workflow, we constructed a novel multimodal music dataset, MMusSet, which contains 33.3k 10-second music clips, with a total duration of 92.5 hours. Each sample in the dataset is a quadruplet consisting of a scene image, story text, music caption, and music segment. To the best of our knowledge, this is the first multimodal music dataset that simultaneously includes images and story texts.

We trained the MusFlow model on the MMusSet dataset and validated its performance through experiments on four tasks: image-to-music, story-to-music, caption-to-music, and multimodal music generation. The experimental results demonstrate that MusFlow is capable of generating high-quality music clips under both unimodal and multimodal conditions.

In summary, this paper makes the following three key contributions:

\begin{itemize}
    \item We are the first to apply the Flow Matching to multimodal music generation, proposing a novel model, MusFlow, which generates music from images and story texts. Experiments demonstrate that MusFlow produces high-quality music conditioned on multimodal inputs.
    \item We introduce an innovative intelligent multimodal music data annotation method, based on the Multi-Agent Workflow. By the collaboration of fine-tuned vision-language models and several pretrained models, we achieved autonomous matching and filtering between image datasets and music datasets.
    \item We constructed a new multimodal music dataset, MMusSet, containing 33k samples. This is the first music dataset that simultaneously includes both images and story texts. We will publicly release this dataset to advance research in multimodal music generation.
\end{itemize}

\begin{figure*}[t]
    \centering
    \includegraphics[height=8cm]{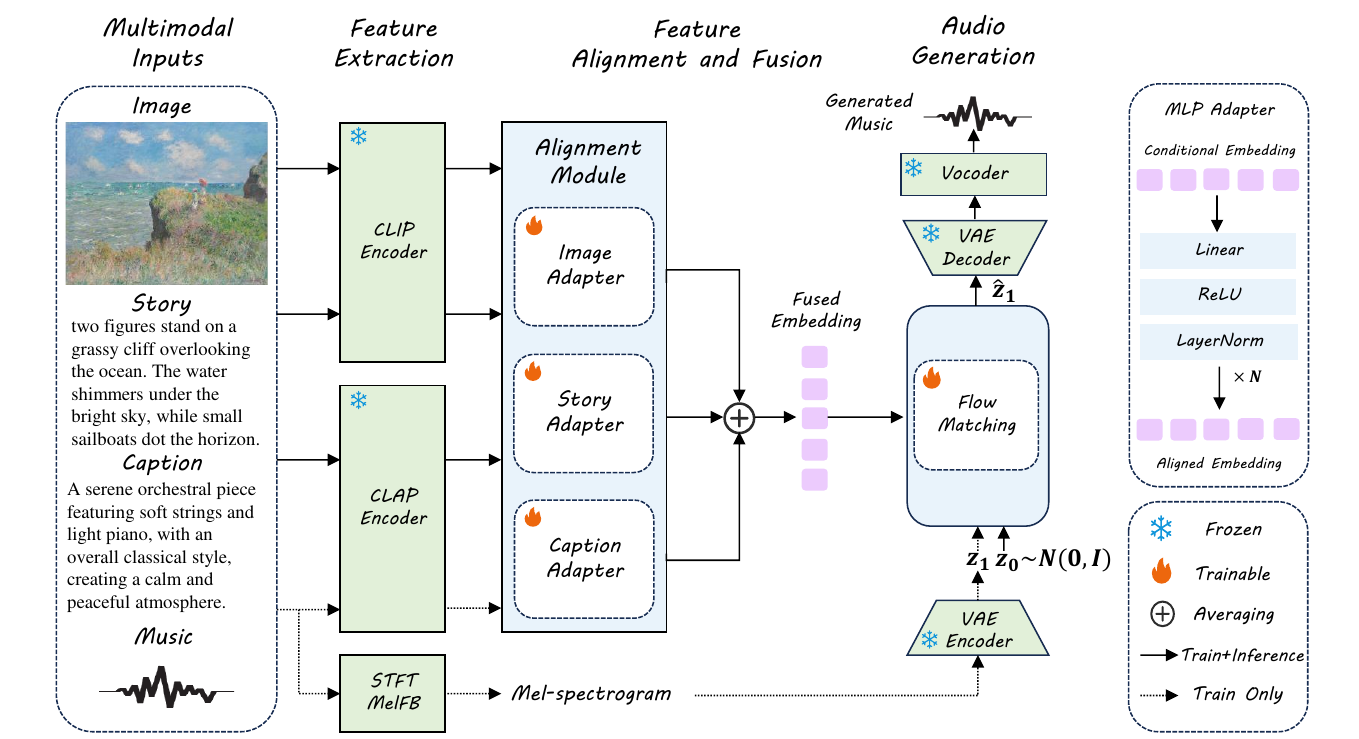}
    \caption{An illustration of our MusFlow framework for multimodal music generation.}
    \label{Figure 2}
\end{figure*}

\section{Related Work}
\subsection{Flow Matching}
Flow Matching (FM) \cite{lipman2023flow, liu2023flow} is a recent advancement in generative modeling that reframes the generation process as learning a continuous-time probability path between a simple prior distribution and a complex data distribution. Unlike SDE-based diffusion models \cite{ho2020denoising, song2021scorebased}, FM leverages continuous normalizing flows (CNFs) \cite{chen2018neural} to construct deterministic trajectories governed by ordinary differential equations (ODEs). This approach aligns with the principles of Optimal Transport (OT) \cite{NIPS2013sinkhorn, 47813}, as FM explicitly minimizes the Wasserstein distance between distributions by learning straight, minimally entropic paths \cite{liu2022rectified, liu2023flow}, bridging OT method and scalable generative modeling. Thanks to its more stable training and faster sampling, FM is widely regarded as a strong successor to traditional diffusion models, particularly in image generation \cite{2024sd3}.

In audio domain, FM has emerged as a powerful framework for tasks. For voice conversion, StableVC \cite{yao2024stablevc} employs conditional FM to disentangle speaker identity and linguistic content, enabling zero-shot style transfer with minimal supervision. Similarly, FM has been applied in text-to-speech (TTS) synthesis: Matcha-TTS \cite{2024Matcha}, CosyVoice \cite{2024CosyVoice} and VoiceFlow \cite{2024voiceflow} leverage FM to achieve real-time synthesis by learning compact, high-fidelity acoustic representations. In music generation, MusicFlow \cite{2024musicflow} adopts cascaded FM to align text descriptions with melodic structures.

In summary, research on FM-based music generation still remains scarce, particularly under multimodal conditions. Our proposed MusFlow addresses this gap by pioneering advancements in this emerging direction.

\subsection{Music Generation}

In recent years, automatic music generation powered by deep learning has made significant progress \cite{zhu2023survey}. Influenced by advancements in image generation, text-to-music has emerged as a widely researched direction. Researchers have adopted powerful architectures from the image generation domain, such as Transformer and diffusion models, to design music generation models conditioned on musical text. MusicGen \cite{NEURIPS2023musicgen} leverages a hierarchical sequence-to-sequence Transformer architecture to generate high-fidelity music from text descriptions. Similarly, MusicLM \cite{agostinelli2023musiclm} explores symbolic music generation, employing a Transformer-based diffusion model that generates discrete sequences in a continuous latent space. Noise2Music \cite{huang2023noise2music} utilizes a two-step diffusion process to refine an intermediate representation of music into high-fidelity audio. Works like AudioLDM2 and MusicLDM focus on latent diffusion method \cite{liu2024audioldm, chen2024musicldm, melechovsky2024mustango}, leveraging a VAE to compress the audio's mel-spectrogram and reconstruct it in the latent space.

Beyond text-based conditioning, some studies have extended music generation to other modalities, such as images and videos. For instance, some researches \cite{2023emotionmusic, hisariya2024bridgingpaintingsmusic} extract emotion-related information from images and use it as a condition to generate music that aligns with the conveyed emotions. Meanwhile, \cite{2021VideoBackgroundMusic, 2024Video2Music, zhuo2023video} focus on video-to-music generation, aiming to produce music that matches the visual content. For multimodal conditions, M$^2$UGen \cite{liu2024mumu} employs a large language model as bridge to convert multimodal inputs into a unified intermediate representation, which serves as a condition for generating music. MeLFusion \cite{2024melfusion} enhances the generation process by using an image as a condition to assist the textual description in producing music.

Unfortunately, most text-to-music models focus primarily on improving the quality of generated music and its alignment with musical descriptions,  overlooking the generalization of input texts. Current multimodal music generation models typically employ large language models to process multimodal inputs and utilize pre-trained models such as MusicGen and AudioLDM2 as the audio generators \cite{liu2024mumu}. This combination requires substantial computational and memory resources, limiting the practicality of these methods. Our proposed model, MusFlow, aims to address these challenges by using simple MLP adapters to handle multimodal information and training a lightweight, end-to-end multimodal music generator based on flow matching.

\section{Methodology}

Multimodal music generation aims to produce music segments that match various multimodal conditions. In this paper, we focus on the task of multimodal music generation conditioned on images, story texts, and music captions. It is important to note that although both story texts and music captions are in text form, they differ significantly: music captions, such as ``a slow-paced classical string composition'', explicitly specify musical elements like instruments and styles, while story texts, such as ``nuns quietly praying in a serene church'', do not contain explicit musical elements. Suppose we have a multimodal music dataset $T=\{I_i, S_i, C_i, M_i\}_{i=1}^{N}$, where $I$ denotes images, $S$ denotes stories, $C$ denotes music captions, and $M$ refers to music segments, the problem we face is how to design and train a multimodal music generation model that, during inference stage, can generate high-quality music segments $M$ based on any combination of $I, S, C$ (with at least one condition provided).

In this section, we talk about how we solve this multimodal music generation task by our proposed MusFlow model. In Sec. \ref{sec 3.1}, we introduce the architecture of the MusFlow model. In Sec. \ref{sec 3.2}, we explain the construction of a Multi-Agent Workflow for collecting the multimodal music dataset MMusSet. In Sec. \ref{sec 3.3}, we describe the training strategies for our model.

\subsection{MusFlow}
\label{sec 3.1}

As shown in Fig. \ref{Figure 2}, given multimodal inputs, the generation process of MusFlow is divided into three stages: Feature Extraction, Feature Alignment and Fusion, and Audio Generation. In this section, we introduce the first two stages in Sec. \ref{sec 3.1.1}, and Audio Generation in Sec. \ref{sec 3.1.2}.

\subsubsection{\textbf{Multimodal Features Extraction, Alignment and Fusion}}
\label{sec 3.1.1}

Given the multimodal inputs $\{I, S, C, M\}$, we first employ pretrained encoders to extract feature embeddings that capture the semantic representations of each input modality. For both the image $I$ and the story text $S$, we use a pretrained CLIP-ViT-B/32 \footnote{\url{https://huggingface.co/openai/clip-vit-base-patch32}} encoder to extract feature embeddings. CLIP \cite{radford2021learning}, which is trained on large-scale image-text paired datasets using contrastive learning \cite{jaiswal2020survey}, can effectively capture the semantic features of both images and texts, and has been widely applied in multimodal vision tasks. Through CLIP encoder $E_{CLIP}$, we can get image embedding $e_{I}$ and story embedding $e_{S}$:
\begin{equation}
    e_{I}=E_{CLIP}(I), \quad e_{S} = E_{CLIP}(S)
\end{equation}

Similarly, for the input music caption $C$ and music segment $M$, we employ a pretrained audio feature extractor, CLAP \cite{laionclap2023}, which is also based on contrastive learning. Specifically, we use a model checkpoint \footnote{\url{https://huggingface.co/lukewys/laion\_clap/blob/main/music\_audioset\_epoch\_15\_esc\_90.14.pt}} that has been pretrained on music datasets to ensure more effective extraction of the semantic features related to music:
\begin{equation}
    e_{C}=E_{CLAP}(C), \quad e_{M} = E_{CLAP}(M)
\end{equation}

To facilitate subsequent feature alignment and fusion, the dimensionality of the feature embeddings extracted from inputs is all 512, i.e. $e_{i} \in \mathbb{R}^{512}, \forall i \in \{I, S, C, M\}$.

Feature alignment is one of the core challenges in multimodal generation. Its goal is to align the feature spaces of different modalities to ensure semantic consistency. In this paper, we explore how to achieve feature alignment, so that during inference, the model can maintain semantic coherence in the generated music, regardless of the modality of conditions. Most previous research has used large language models to unify the semantic features across different modalities \cite{liu2024audioldm, liu2024mumu}. While this approach achieves good alignment, it comes with significant resource consumption. In this work, we aim to design a simpler and more efficient alignment method. In real-world scenarios, humans can naturally associate scenes with fitting music through artistic and emotional perception. In our generation task, the model is expected to exhibit a similar ability by extracting musical features from multimodal conditions. During the feature extraction stage, we use the CLAP encoder to obtain feature embeddings of real music, which serve as a latent representation of musical semantics. Therefore, we model this ``music association ability'' as effective mappings from the conditional embeddings, $e_{I},e_{S}$ and $e_{C}$, to the real music embeddings $e_{M}$. Through these mappings, we can embed the conditional embeddings into the CLAP audio feature space, achieving feature alignment.

Inspired by the alignment practices in vision language models \cite{NEURIPS2023llava}, we employ MLP adapters to learn these mapping relationships. As shown in Fig. \ref{Figure 2}, we set up separate MLP adapters for each of the three conditions to learn the mappings from their respective feature spaces to the CLAP audio feature space. We use the MSE as the alignment loss, with the loss function formulated as follows:
\begin{equation}
    \mathcal{L_{A}} = ||e_{M} - \frac{1}{n}\sum_{i}MLP_{i}(e_{i}) ||^{2}, \quad i\in \{I, S, C\}
    \label{Equation 3}
\end{equation}
where $n$ represents the number of input conditions. By minimizing $\mathcal{L_{A}}$, the MLP adapters learn effective mappings that transform the conditional embeddings, $e_{I},e_{S}$ and $e_{C}$, into the target music CLAP embeddings $e_{M}$, which contain musical semantics. These transformed embeddings are then averaged to obtain a unified aligned embedding $e_{f}$, which serves as the final conditional embedding in generation process.

Previous work \cite{chen2024musicldm} has observed that directly using the music CLAP embeddings as conditions when training generative models leads to better generation performance. This is understandable, as the CLAP embeddings of the music itself contain richer musical semantic features, which can better guide the generation process. We also verify this phenomenon in our experiments. Our feature alignment method enables the direct mapping of conditional embeddings into the real music CLAP feature space, thereby more effectively guiding the generation process. We designed experiments to verify the effectiveness of this alignment method, and the details can be found in Sec. \ref{sec 4.3}.

\begin{figure*}[t]
    \centering
    \includegraphics[height=6.5cm]{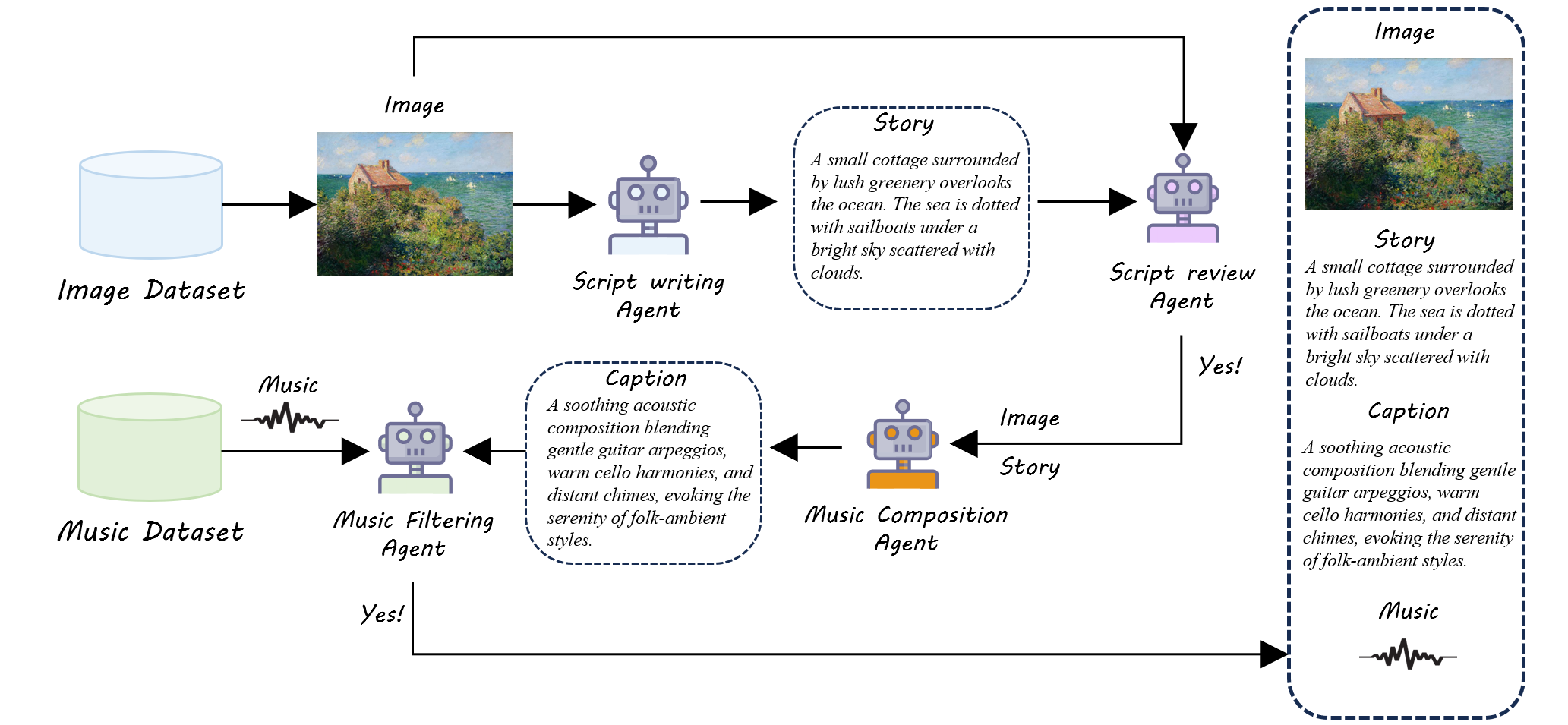}
    \caption{An illustration of our proposed Multi-Agent Workflow for dataset creation.}
    \label{Fig. 3}
\end{figure*}

\subsubsection{\textbf{Audio Generation via Conditional Flow Matching}}
\label{sec 3.1.2}

In this section, we first introduce the foundational framework of conditional flow matching, then we define the training objective of the flow matching module in the audio generation part.

Flow matching \cite{lipman2023flow} is a method used to fit the time-dependent probability path between a prior and a target distribution. It is derived from Continuous Normalizing Flows (CNFs) \cite{chen2018neural} but offers a more efficient simulation-free training approach. CNFs construct a probability path between the prior distribution $p_{0}(x)$ (typically Gaussian) and the data distribution $p_{1}(x)$ through a time-dependent vector field $v_{t}(x): [0,1] \times \mathbb{R}^{d} \rightarrow \mathbb{R}^{d}$. The flow $\phi_{t}(x): [0,1] \times \mathbb{R}^{d} \rightarrow \mathbb{R}^{d}$ controlled by $v_{t}(x)$ is defined via the ODE:
\begin{equation}
    \frac{d}{dt}\phi_{t}(x) = v_{t}(\phi_{t}(x)); \quad \phi_{0}(x)=x
\end{equation}
and the probability density $p_{t}(x)$ evolves according to the change of variables formula for $p_{t}(x)=p_{0}(\phi_{t}^{-1}(x))\det{\left[ \frac{\partial\phi_{t}^{-1}}{\partial{x}}(x) \right]}$.


Flow Matching aims to solve the probability path by fitting the true vector filed $v_{t}(x)$ through a learnable neural network $v_{\theta}(x,t)$:
\begin{equation}
    \mathcal{L}_{FM} = \mathbb{E}_{t, p_{t}(x)} || v_{\theta}(x,t) - v_{t}(x) ||^{2}
\end{equation}
while the true vector field $v_{t}(x)$ is generally intractable,  Conditional Flow Matching (CFM) \cite{lipman2023flow} circumvents this by factorizing the problem using conditional probability paths $p_{t}(x|z)$:
\begin{equation}
    \mathcal{L}_{CFM} = \mathbb{E}_{t, q(z), p_{t}(x|z)} || v_{\theta}(x,t) - v_{t}(x|z) ||^2 \label{Equation 7}
\end{equation}
if we can efficiently sample from $q(z)$ and $p_{t}(x|z)$, and calculate $v_{t}(x|z)$, we can use $\mathcal{L}_{CFM}$ to approximate the marginal vector field $v_{t}(x)$, as $v_{t}(x):=\mathbb{E}_{q(z)}\frac{v_{t}(x|z)p_{t}(x|z)}{p_{t}(x)}$, and it has also been improved that $\nabla_{\theta}\mathcal{L}_{FM}=\nabla_{\theta}\mathcal{L}_{CFM}$, guaranteeing the optimization equivalence between $\mathcal{L}_{FM}$ and $\mathcal{L}_{CFM}$. 

There are various settings of $q(z)$ and $p(x|z)$, leading to different types of CFM, such as FM-OT, FM-Diffusion \cite{lipman2023flow}, OT-CFM, SB-CFM \cite{tong2024improving} and Rectified Flow \cite{liu2022rectified, albergo2023building}. Building on the successful application of the FM-OT method in the field of audio generation \cite{yao2024stablevc, 2024Matcha, 2024musicflow}, we follow previous works and adopt the FM-OT formulation of CFM. FM-OT identifies the condition $z$ with data distribution $p_{1}(x)$, and considers a Gaussian distributions for $p_{t}(x|x_{1})=\mathcal{N}(\mu_{t}(x_1), \sigma_{t}(x_1))$, where $\mu_{t}(x_1)=tx_1$ and $\sigma_{t}^{2}(x_1)=1-(1-\sigma_{min})t$. The corresponding training objective can be written as
\begin{equation}
    \mathcal{L}_{FM-OT}=\mathbb{E}_{t,p(x_{1}),p_{t}(x|x_{1})}||v_{\theta}(x,t)-(x_1-(1-\sigma_{min})x_0)||^2
    \label{Equation 7}
\end{equation}

Under FM-OT, the conditional flow $\phi_{t}(x)$ is in fact the Optimal Transport displacement map \cite{MCCANN1997153} between the two Guassian $p_{0}(x|x_1)$ and $p_{1}(x|x_1)$, resulting in a straight-line probability transport trajectory with constant speed. \cite{lipman2023flow}.

In the audio generation part of our method, we first apply the Short-Time Fourier Transform (STFT) to obtain the Mel-spectrogram $M_{mel} \in \mathbb{R}^{T \times F}$ of the real music $M$, where $T$ denotes the time domain and $F$ denotes the frequency domain. Inspired by previous works \cite{liu2024audioldm, agostinelli2023musiclm, melechovsky2024mustango}, we use the pretrained VAE in MusicLDM \cite{agostinelli2023musiclm} to obtain a compressed latent representation $Z_{1} = V_{Encoder}(M_{mel}) \in \mathbb{R}^{C \times H \times W}$. The probability distribution of latent variable $Z_{1}$ is the target distribution that the flow matching model aims to generate. During training, by initializing standard Gaussian $Z_{0} \sim \mathcal{N}(0, I)$ as the original distribution, we can get the training objective of our flow matching module according to Eq. \eqref{Equation 7}:
\begin{equation}
    \mathcal{L_{G}} = \mathbb{E}_{t, q(z_1), p(z|z_1)}||v_{\theta}(z,t,e_{f})-(z_{1}-(1-\sigma_{min})z_{0})||^2
    \label{Equation 8}
\end{equation}\
where $e_{f}$ represents the fused conditional embedding. The backbone model for flow matching adopts the Transformer-Unet architecture \cite{rombach2022high}, which is widely used in image generation tasks. By minimizing $\mathcal{L_{G}}$, we can estimate the true vector field and use numerical methods to obtain the probability path, allowing us to generate new sample $\hat{Z}_{1}$ from the target distribution. During inference, the sampled $\hat{Z}_{1}$ is decoded into the predicted Mel-spectrogram $\hat{M}_{mel}$ using the VAE Decoder. Finally, through pretrained Hifi-GAN vocoder \cite{2020hifigan, agostinelli2023musiclm}, we can transform $\hat{M}_{mel}$ into music waveform.

In summary, by effectively training the alignment adapters and flow matching module in Fig. \ref{Figure 2}, we are able to accurately map multimodal conditions into the CLAP audio feature space, and sample high-quality music audio based on the fused conditional embedding. To achieve better generation performance, we employ a multi-stage training approach, which will be detailed in Sec. \ref{sec 3.3}.

\subsection{Multi-Agent Data Annotation Workflow}
\label{sec 3.2}
One of the core challenges currently faced by multimodal music generation is the scarcity of datasets. Existing datasets are mostly paired collections of music descriptions and audios, lacking the diversity of modality pairings needed for further research in this field. To address the scarcity of multimodal data while avoiding the high time and labor costs, in this section, we introduce a data annotation workflow based on Multi-Agent framework. Using this workflow, we create a new multimodal music dataset, MMusSet.

The Multi-Agent Workflow \cite{hong2024metagpt} refers to design a pattern where multiple autonomous agents  divide tasks and collaborate to complete complex assignments according to a specific workflow. In this paper, we need to define the tasks for each agent and design a tailored workflow that enables autonomous multimodal music data annotation. Since the production of music for films and games involves multimodal information, such as scene art, scripts, and background music, we simulate a similar music production workflow as shown in Fig. \ref{Fig. 3}. We design four agents in this workflow.

\textbf{Script Writing Agent}. Its task is to write a descriptive text for each image in the provided image dataset, serving as the story text that semantically matches the corresponding scene. This is essentially an image captioning task, for which we selected the BLIP model \cite{li2022blip} as the underlying model for this agent, as it can efficiently handle image captioning task.

\textbf{Script Review Agent}. This agent evaluates the image-story pairs from the previous stage to ensure semantic relevance and filter out any low-quality data generated by the Script Writing Agent. For this, we use the CLIP model \cite{radford2021learning}. By calculating the CLIP score \cite{2021clipscore}, we can accurately assess the relevance between the image and the story. Through setting a similarity threshold, we can filter out low-relevance samples.

\textbf{Music Composition Agent}. This agent's task is to compose a matching background music caption based on the image and story text, similar to how composers create background music based on scene art and scripts. This task requires the model to have a certain level of music understanding. We chose Qwen2-VL-7B \cite{wang2024qwen2} as the underlying model and fine-tuned it using a dataset of image-story-caption triplets, which will be introduced in detail later.

\textbf{Music Filtering Agent}: This agent's task is to select matching music samples from a public music dataset based on the music caption provided by the Music Composition Agent. The agent needs to assess the relevance between the music and the descriptive text, so we use the CLAP model \cite{laionclap2023} for this task. By calculating the CLAP score between the caption and all music clips in the provided dataset, the agent selects the highest-scoring clip, i.e., the most relevant one. It also checks if the similarity exceeds a predefined threshold to determine whether a successful match has been made.

Through the workflow composed of the four agents mentioned above, we can achieve automatic matching and filtering between a music dataset and a image dataset, autonomously constructing a multimodal music dataset that includes images, story texts, music captions, and music audios.

In previous work, there are several datasets which have already matched music with images, including Music2Image \footnote{\url{https://huggingface.co/datasets/Monke64/Music2Image}}, extended-MusicCaps, MelBench \cite{2024melfusion} and MUImage \cite{liu2024mumu}. Using the Script Writing Agent, we generate story texts for these datasets, which are then filtered through the Script Review Agent. Ultimately, we obtain 10k pretraining data samples. With this pretraining data, we fine-tune the Qwen2-VL-7B model using QLoRA \cite{dettmers2023qlora}, enabling it to generate more accurate music captions based on images and stories. The fine-tuning effects are demonstrated in Sec. \ref{sec 4.4}. After fine-tuning, we use the ArtGraph \cite{CASTELLANO2022108859} as the image dataset and FMA-Medium \cite{defferrard2017fmadatasetmusicanalysis} as the music dataset to create a new multimodal music dataset, ArtMusic, based on the workflow. Combining the processed pretraining data and ArtMusic, we constructed a new multimodal music dataset called MMusSet, which contains 33.3k samples. Using this dataset, we train the MusFlow model for multimodal music generation.

\subsection{Training Strategies}
\label{sec 3.3}

In this section, we introduce the training strategies for MusFlow. To effectively enhance the music generation quality, we adopt a multi-stage training approach. The entire training process is divided into three phases: Alignment Training, Generation Training, and Joint Training. Additionally, to simulate different combinations of multimodal inputs, we employ a random condition masking method during training to improve the model's robustness. Next, we will provide a detailed explanation of these strategies.

\textbf{Alignment Training}. In the first phase, we train the MLP adapters used for feature alignment independently. By minimizing $\mathcal{L_{A}}$ in Eq. \eqref{Equation 3}, we learn effective mappings from multimodal conditional embeddings to the CLAP audio feature space, obtaining fused conditional embeddings in a unified space, which facilitates the subsequent generation step. 

\textbf{Generation Training}. In the second phase, we freeze the MLP adapters and train the flow matching module independently. The primary goal at this stage is to enhance the quality of audio generation by optimizing $\mathcal{L_{G}}$ in Eq. \eqref{Equation 8}. A larger number of training epochs is required at this stage to achieve better results.

\textbf{Joint Training}. In the third phase, we train both the MLP adapters and the flow matching module simultaneously to eliminate the accumulated errors from separate training. Our loss function is the sum of the generation loss and the alignment loss:
\begin{equation}
    \mathcal{L_{J}} = \mathcal{L_{G}}+\lambda\mathcal{L_{A}}
\end{equation}
where $\lambda$ is a parameter that controls the importance of the alignment loss. Through this phase of training, we can further improve the model's generation quality and condition relevance.

\textbf{Random Condition Mask}. In practical applications, users may provide a subset of the input conditions. Only relying on full-condition training can reduce the model’s adaptability to incomplete inputs. To address this, we apply random masking with equal probability to the input conditions in each batch during training, while ensuring that at least one condition remains. This yields seven possible masking combinations, allowing the same sample to be trained under diverse input scenarios across epochs, thereby improving both data efficiency and model generalization.

\section{Experiments}

\begin{table*}[htbp]
\centering
\small 
\setlength{\tabcolsep}{4pt} 
\def\arraystretch{1.2}
\caption{\textbf{Comparison of Models for Music Generation (N/A indicates that the model is unable to handle this type of task.)}.}
\label{tab 1}
\begin{tabular}{c|ccc|ccc|ccc|ccc}
\hline \hline
\multirow{2}{*}{\textbf{Model}} & \multicolumn{3}{c|}{\textbf{Caption-to-Music}} & \multicolumn{3}{c|}{\textbf{Story-to-Music}} & \multicolumn{3}{c|}{\textbf{Image-to-Music}} & \multicolumn{3}{c}{\textbf{Multimodal Generation}} \\ 
\cline{2-13} 
 & \textit{\textbf{FAD$\downarrow$}} & \textit{\textbf{KL$\downarrow$}} & \textit{\textbf{CLAP$_{score}$$\uparrow$}} & \textit{\textbf{FAD$\downarrow$}} & \textit{\textbf{KL$\downarrow$}} & \textit{\textbf{CLAP$_{score}$$\uparrow$}} & \textit{\textbf{FAD$\downarrow$}} & \textit{\textbf{KL$\downarrow$}} & \textit{\textbf{IB$_{score}$$\uparrow$}} & \textit{\textbf{FAD$\downarrow$}} & \textit{\textbf{CLAP$_{score}$$\uparrow$}} & \textit{\textbf{IB$_{score}$$\uparrow$}} \\ 
\hline \hline
\textit{MusicGen} & 5.36 & 3.7 & 0.20 & 7.72 & 4.04 & 0.18 & N/A & N/A & N/A & N/A & N/A & N/A \\ 
\textit{AudioLDM2-Music} & 4.24 & 2.86 & 0.28 & 7.34 & 3.33 & 0.22 & N/A & N/A & N/A & N/A & N/A & N/A \\ 
\textit{MusicLDM} & 3.53 & 2.59 & \textbf{0.34} & 3.98 & 3.16 & 0.29 & N/A & N/A & N/A & N/A & N/A & N/A \\ 
\textit{CoDi} & 5.23 & 3.54 & 0.21 & 8.45 & 4.26 & 0.19 & 7.34 & 3.86 & 0.60 & 6.89 & 0.21 & 0.62 \\ 
\textit{M$^{2}$UGen} & 4.84 & 2.76 & 0.24 & 5.83 & 3.25 & 0.27 & 6.24 & 3.75 & 0.71 & 6.36 & 0.25 & 0.68 \\
\midrule
\textit{\textbf{MusFlow}} & \textbf{3.41} & \textbf{2.56} & 0.32 & \textbf{3.56} & \textbf{2.78} & \textbf{0.31} & \textbf{3.66} & \textbf{2.79} & \textbf{0.82} & \textbf{3.36} & \textbf{0.33} & \textbf{0.83} \\ 
\hline \hline
\end{tabular}
\end{table*}

In this section, we conduct a comprehensive evaluation of the proposed MusFlow model across four tasks: Image-to-Music, Story-to-Music, Caption-to-Music, and Multimodal Music Generation. Additionally, we design ablation studies to verify the effectiveness of the proposed feature alignment method. Furthermore, we assess the impact of fine-tuning Qwen2-VL in the data creation workflow to ensure the construction of a higher-quality dataset.

\subsection{Experimental Setup}
\subsubsection{\textbf{Dataset}} 
We train the MusFlow model using MMusSet, a newly constructed multimodal music dataset proposed in this paper. MMusSet is built upon the multi-agent music data annotation workflow, leveraging publicly available datasets such as Music2Image, extended-MusicCaps, MelBench, MUImage, ArtGraph, and FMA-Medium. The detailed construction methodology can be found in Sec. \ref{sec 3.2}. The specific statistics of MMusSet are presented in the Tab. \ref{tab 2}.

\begin{table}
  \small
  \caption{Basic Information of MMusSet}
  \centering
  \label{tab 2}
  \begin{tabular}{ccccc}
    \toprule
    \textit{\textbf{Dataset}} & \textit{\textbf{Num (k)}} & \textit{\textbf{Dur (h)}} \\
    \midrule
    \textit{Music2Image} & 1.50 & 4.17 \\
    \textit{MusicCaps} & 4.54 & 12.60  \\
    \textit{MelBench} & 0.98 & 2.69 \\
    \textit{MUImage} & 3.83 & 10.64 \\
    \textit{ArtMusic} & 22.46 & 62.39 \\
    \midrule
    \textit{\textbf{MMusSet (Total)}} & \textbf{33.31} & \textbf{92.49}\\
  \bottomrule
\end{tabular}
\end{table}

Notably, in the Multi-Agent Workflow, we set the filtering thresholds to 0.3 for the Script Review Agent and Music Filtering Agent. For the experiments, we use extended-MusicCaps, MelBench, MUImage, and 90\% of ArtMusic for training, while Music2Image and 10\% of ArtMusic are used for evaluation.

\subsubsection{\textbf{Hyperparameters}}

We train MusFlow on a NVIDIA RTX 4090 GPU. During training, we use the AdamW optimizer with a learning rate of 1e-4 and set the batch size to 24. For the MLP adapters in the Alignment Module, we configure the network with 4 layers. The training process consists of three stages: alignment training, generation training, and joint training, with 50, 250, and 100 epochs, respectively. For the Transformer-Unet backbone in the flow matching module, we largely follow the parameter settings of MusicLDM \cite{chen2024musicldm}. For fine-tuning Qwen2-VL, we apply QLoRA with a rank of 8 and train for 20 epochs.

\subsubsection{\textbf{Evaluation Metrics}}

We perform both objective evaluation and human subjective evaluation. For objective evaluation, we select Frechet Audio Distance (FAD), Kullback–Leibler (KL) Divergence \cite{liu2024audioldm}, CLAP Score, and ImageBind Score as evaluation metrics. Specifically, FAD and KL measure the statistical similarity and perceptual quality between the generated and real audio samples, while CLAP Score and ImageBind Score \cite{2023imagebind} assess the semantic alignment between the generated audio and the input text and image, respectively. For subjective evaluation, we use Overall Quality (OVL) and Relevance to Input Condition (REL) \cite{liu2024audioldm} as evaluation metrics. These metrics assess the generation quality and condition adherence of the generated audio, respectively.

\begin{figure*}[t]
    \centering
    \includegraphics[height=4cm]{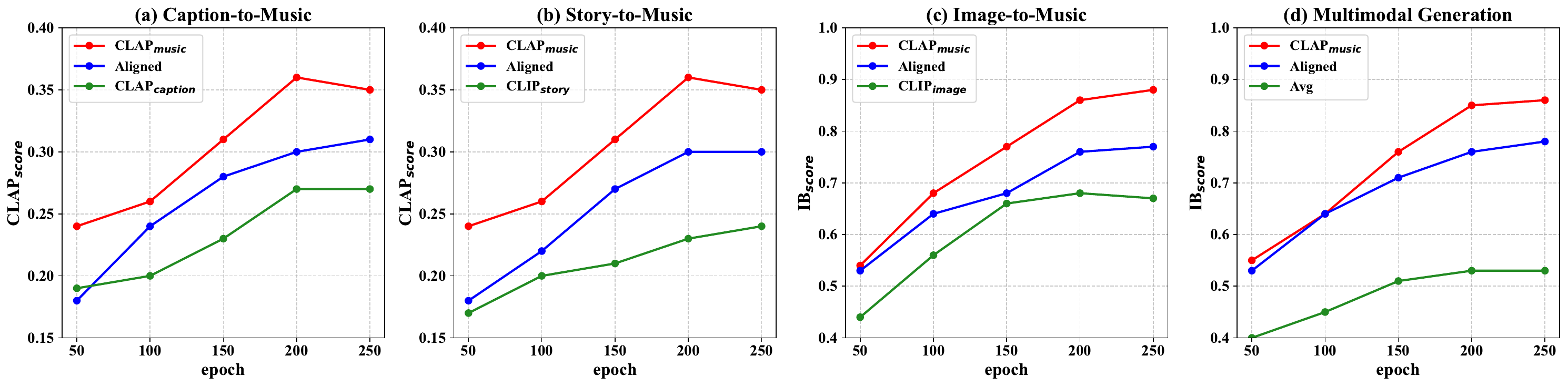}
    \caption{Ablation study results of the Alignment Module. CLAP$_{music}$ refers to using the target audio's CLAP embedding as conditioning during training; CLAP$_{caption}$ and CLIP$_{story}$ represent directly using the original features as conditioning; Avg denotes using the average of the original features as conditioning across multimodal task; Aligned refers to the proposed method, where the Alignment Module is applied to obtain the conditional embedding.}
    \label{Fig. 4}
\end{figure*}

\subsection{MusFlow Evaluation}
\subsubsection{\textbf{Caption-to-Music Generation}}

We first evaluate the performance of MusFlow on the traditional text-to-music generation task based on music descriptions. The selected baseline models include MusicGen \cite{NEURIPS2023musicgen}, AudioLDM2 \cite{liu2024audioldm}, MusicLDM \cite{chen2024musicldm}, CoDi \cite{tang2023anytoany}, and M$^{2}$UGen \cite{liu2024mumu}, which are widely used in the field of music generation. The evaluation results on the validation set are presented in Tab. \ref{tab 1}. The MusFlow model achieves the lowest FAD of 3.41 and the lowest KL of 2.56, indicating that MusFlow attains the highest audio generation quality among all evaluated models. Additionally, MusFlow achieves a CLAP Score of 0.32, which is slightly lower than MusicLDM's 0.34, but still demonstrates a high degree of semantic relevance in generated audio.

\subsubsection{\textbf{Story-to-Music Generation}}

Unlike the traditional text-to-music generation task based on music descriptions, Story-to-Music Generation aims to generate background music from story texts that do not explicitly contain musical attributes. This task requires the model to comprehend the underlying musical characteristics in the narrative. We use the same baseline models as in the Caption-to-Music task for comparative experiments. The results, presented in Tab. \ref{tab 1}, show that MusFlow achieves the best performance across all three metrics: FAD, KL, and CLAP Score. This demonstrates that the model effectively captures and translates the implicit musical attributes within story texts into coherent background music.

\subsubsection{\textbf{Image-to-Music Generation}}

Image-to-Music Generation aims to generate music that align with the style, emotion, and atmosphere of a given image, making it a crucial direction in multimodal music generation. For this task, we select CoDi and M$^{2}$UGen as baseline models. The experimental results, presented in Tab. \ref{tab 1}, show that MusFlow achieves the lowest FAD of 3.66 and KL of 2.79, indicating the highest audio generation quality among the evaluated models. Additionally, MusFlow surpasses the baseline models in IB Score, reaching 0.82, which demonstrates that the generated music achieves the best semantic alignment with the input image.

\subsubsection{\textbf{Multimodal Music Generation}}

Multimodal music generation requires models to generate music clips that align with multimodal input conditions, demanding a strong understanding of musical attributes. This makes it a highly challenging research direction. In this study, our input conditions include music captions, story texts, and images. To evaluate the model's robustness in handling arbitrary multimodal input combinations, we adopt the random condition masking strategy introduced in Sec. \ref{sec 3.3}, simulating real-world cases where any subset of modalities may be available. For comparison, we select M$^{2}$UGen and CoDi as the baseline model. As shown in tab \ref{tab 1}, MusFlow outperforms M$^{2}$UGen and CoDi across three key metrics: FAD, CLAP Score, and IB Score. This indicates that our model achieves higher audio generation quality and better semantic alignment, demonstrating superior performance in multimodal music generation.

\subsubsection{\textbf{Subjective Evaluation}}

\begin{table}
  \small
  \caption{Subjective evaluation results for different models.}
  \centering
  \label{tab 3}
  \begin{tabular}{ccccc}
    \toprule
    \textit{\textbf{Model}} & \textit{\textbf{OVL$\uparrow$}} & \textit{\textbf{REL$\uparrow$}} \\
    \midrule
    \textit{MusicGen} & 64.82 & 62.11 \\
    \textit{AudioLDM2-Music} & 69.55 & 72.39  \\
    \textit{MusicLDM} & 76.45 & 73.25 \\
    \textit{CoDi} & 62.37 & 63.49 \\
    \textit{M$^2$UGen} & 70.54 & 69.23 \\
    \midrule
    \textit{\textbf{MusFlow}} & \textbf{78.25} & \textbf{74.12}\\
  \bottomrule
\end{tabular}
\end{table}

In the previous evaluation tasks, we randomly selected 50 generated samples from each model for human evaluation. A total of 20 participants were invited to assess the generated samples based on subjective metrics. The evaluation results are presented in Tab. \ref{tab 3}. The results show that MusFlow achieves the highest scores in both OVL and REL. This indicates that the samples generated by MusFlow are more aesthetically pleasing and better aligned with human subjective preferences compared to other models.

\subsection{Feature Aligenment Evaluation}
\label{sec 4.3}

To validate the effectiveness of the Alignment Module in Fig. \ref{Figure 2}, we conduct ablation experiments. The MLP adapters within the Alignment Module are designed to map multimodal conditioning information into the CLAP feature space of the target audio. To assess their impact, we train additional models in the Audio Generation stage: one using the CLAP embedding of the target audio as direct conditioning and another without the Alignment Module, directly using original feature embedding extracted by CLIP and CLAP encoder.

During training, we evaluated these three models on the validation set every 50 epochs, comparing the generated audios based on semantic relevance metrics, CLAP$_{score}$ and IB$_{score}$. The evaluation results are shown in Fig. \ref{Fig. 4}. Across all evaluation tasks, we observed that directly using the target audio’s CLAP embedding as a conditioning input led to superior generation quality. This is expected, as these embeddings inherently contain the target audio's features. This phenomenon has also been noted in previous work \cite{chen2024musicldm}. The results demonstrate that, compared to directly using raw feature embeddings, incorporating the MLP adapters in the Alignment Module brings our model’s generation performance closer to that of the model conditioned directly on the target audio’s CLAP embedding, enabling the generation of music with higher semantic alignment. This indicates that the Alignment Module enables our model to extract key musical features from multimodal conditioning information, thereby providing better guidance for the generation process.

\subsection{Fine-tuned Qwen2-VL Evaluation}
\label{sec 4.4}

The quality of the MMusSet dataset constructed by our Multi-Agent Workflow largely depends on whether Qwen2-VL can generate music captions that accurately align with the given images and story texts. To evaluate the impact of fine-tuning on Qwen2-VL’s performance, we conduct a set of experiments.

Using the QLoRA method, we fine-tune Qwen2-VL on MusicCaps, MelBench, and MUImage, and validate the fine-tuned model on the Music2Image dataset. We adopt CLAP$_{score}$ and BART$_{score}$ \cite{2021bartscore} as evaluation metrics. CLAP$_{score}$ measures the alignment between the generated caption and the target music, while BART$_{score}$, computed by a pretrained language model, MiniLM-V2 \footnote{\url{https://huggingface.co/sentence-transformers/all-MiniLM-L6-v2}} \cite{2021minilmv2}, assesses the textual similarity between the generated caption and the target caption. The evaluation results are shown in Tab. \ref{tab 4}.

\begin{table}
  \small
  \caption{Fine-tuning results for Qwen2-VL.}
  \centering
  \label{tab 4}
  \begin{tabular}{ccc}
    \toprule
    \textit{\textbf{Model}} & \textit{\textbf{CLAP$_{score}$$\uparrow$}} & \textit{\textbf{BART$_{score}$$\uparrow$}} \\
    \midrule
    \textit{base} & 0.15 & 0.50 \\
    \textit{fine-tuned (rank 4)} & 0.18 & 0.54 \\
    \textit{fine-tuned (rank 8)} & 0.24 & 0.63 \\
    \bottomrule
\end{tabular}
\end{table}

The results indicate that fine-tuned Qwen2-VL generates captions with improved similarity to both the target captions and the corresponding audios. Furthermore, increasing the rank of the low-rank matrices in QLoRA further enhances fine-tuning performance, demonstrating that fine-tuning contributes to better data generation. Combined with the other agents in the Multi-Agent Workflow, the fine-tuned Qwen2-VL ensures the construction of a higher-quality dataset.

\section{Limitations}

In this section, we discuss the limitations of our study. Firstly, MusFlow is trained on music data with a sampling rate of 16 kHz, whereas most standard music productions use 44.1 kHz. This limitation primarily arises from the target sampling rate constraint of the pretrained HiFi-GAN model. Secondly, due to computational resource constraints, we do not incorporate the video modality when constructing our multimodal dataset. However, video-based music generation is an important research direction that warrants further exploration. Thirdly, owing to GPU memory limitations, we are unable to fine-tune Qwen2-VL using QLoRA with a rank exceeding 8, which may have hindered the model’s ability to generate higher-quality datasets.

\section{Conclusion}

Multimodal music generation aims to generate music clips that align with multimodal conditions while adhering to human aesthetic preferences. In this paper, we propose MusFlow, a novel multimodal music generation model capable of generating music clips based on images, story texts, and musical captions. MusFlow employs MLP adapters for feature alignment and utilizes the conditional flow matching for audio generation. Experimental results demonstrate that MusFlow achieves strong performance across multiple tasks. Furthermore, we introduce MMusSet, a novel multimodal music dataset constructed within the Multi-Agent Workflow framework. We believe that MMusSet will facilitate future research in the field of multimodal music generation.

\bibliographystyle{ACM-Reference-Format}
\bibliography{reference}

\end{document}